\documentclass[a4paper,fleqn,usenatbib]{mnras}

\usepackage{newtxtext,newtxmath}

\usepackage[T1]{fontenc}
\usepackage{ae,aecompl}

\usepackage[normalem]{ulem}

\usepackage{graphicx}	
\usepackage{amsmath}	
\usepackage{amssymb}	

\defcitealias{Stassun2016}{ST16}

\title[Testing Asteroseismic Radii]{Testing Asteroseismic Radii of Dwarfs and Subgiants
                                    with {\it Kepler} and {\it Gaia}}

\author[C. L. Sahlholdt et al.]{
C. L. Sahlholdt$^{1,2}$\thanks{E-mail: sahlholdt@astro.lu.se},
V. Silva Aguirre$^{1}$,
L. Casagrande$^{3}$,
J. R. Mosumgaard$^{1}$,\newauthor
M. Bojsen-Hansen$^{1}$
\\
$^{1}$Stellar Astrophysics Centre, Department of Physics and Astronomy, Aarhus University,
      Ny Munkegade 120, DK-8000 Aarhus C, Denmark\\
$^{2}$Lund Observatory, Department of Astronomy and Theoretical Physics, Box 43,
      SE-221 00 Lund, Sweden \\
$^{3}$Research School of Astronomy \& Astrophysics, Mount Stromlo Observatory,
      The Australian National University, ACT 2611, Australia
}

\date{Accepted XXX. Received YYY; in original form ZZZ}

\pubyear{2017}

\begin{document}
\label{firstpage}
\pagerange{\pageref{firstpage}--\pageref{lastpage}}
\maketitle

\begin{abstract}
We test asteroseismic radii of {\it Kepler} main-sequence and subgiant stars by deriving
their parallaxes which are compared with those of the first {\it Gaia} data release.
We compute radii based on the asteroseismic scaling relations as well as by fitting
observed oscillation frequencies to stellar models for a subset of the sample, and test
the impact of using effective temperatures from either spectroscopy or the infrared flux
method.
An offset of 3\%, showing no dependency on any stellar parameters, is found between
seismic parallaxes derived from frequency modelling and those from {\it Gaia}.
For parallaxes based on radii from the scaling relations, a smaller offset is
found on average; however, the offset becomes temperature dependent which we interpret
as problems with the scaling relations at high stellar temperatures.
Using the hotter infrared flux method temperature scale, there is no indication that
radii from the scaling relations are inaccurate by more than about 5\%.
Taking the radii and masses from the modelling of individual frequencies as reference
values, we seek to correct the scaling relations for the observed temperature trend.
This analysis indicates that the scaling relations systematically overestimate radii
and masses at high temperatures, and that they are accurate to within 5\% in radius and
13\% in mass for main-sequence stars with temperatures below 6400~K.
However, further analysis is required to test the validity of the corrections on a
star-by-star basis and for more evolved stars.
\end{abstract}




\section{Introduction}
With the advent of remarkably precise space-based photometry in recent years, the field
of asteroseismology has seen great progress as a means of determining precise stellar
parameters.
For a star showing solar-like oscillations, two global parameters can be determined from
the power spectrum, namely the mean large frequency separation,
$\langle\Delta\nu\rangle$\footnote{Since we use the mean value throughout this paper we
drop the brackets and simply write $\Delta\nu$.}, and the frequency of
maximum oscillation power, $\nu_{\mathrm{max}}$.
These parameters follow a set of approximate {\it scaling relations}, tying them to the
mass and radius of the star, which makes it straightforward to estimate properties of any
star with detected solar-like oscillations.
Therefore, asteroseismology holds great potential for determining precise stellar
parameters, and it is important to verify that the results are also accurate.

Direct measurements of radii and masses are challenging to obtain which makes it
difficult to perform large-scale tests of the scaling relations.
Still, empirical tests have been carried out for smaller samples, and the scaling
relation for the stellar radius has been shown to be accurate to within 4\% for
main-sequence and subgiant stars based on comparisons with results from interferometry
\citep{Huber2012,White2013}.
The scaling relation for the mass can be tested with eclipsing binaries in which one
of the components shows solar-like oscillations. \citet{Gaulme2016} compared both masses
and radii based on eclipse analyses with those from the scaling relations for a sample of
10 red giants.
They found that the scaling relations overestimate radii and masses by
about 5\% and 15\% respectively, emphasising the need for further investigation of the accuracy of
the scaling relations.
In order to test the results of asteroseismology for larger stellar samples, less direct
comparisons must be employed.
\citet{SilvaAguirre2012} used the scaling relations in combination with broadband
photometry to derive stellar distances which they compared to {\it Hipparcos}
parallaxes.
Based on their sample of 22 main-sequence stars, they found the asteroseismic distances,
and thereby also the radii, to be accurate to within 5\%.

With the recent first data release from the {\it Gaia} mission
(GDR1; \citealt{collaboration2016}), a new opportunity to test asteroseismic radii has
arisen.
The {\it Gaia} data have significantly increased the number of solar-like oscillators with
precise parallax measurements and allow for new observational constraints to be put on
the accuracy of asteroseismic radii.
A first comparison between seismic and {\it Gaia} parallaxes was carried out by
\citet{DeRidder2016} who found good agreement for 22 dwarfs and subgiants; however,
for 938 red giants they could reject the 1:1 relation at the 95\% confidence limit.
\citet{Huber2017} used the {\it Gaia} parallaxes to test the scaling relations for a sample of
2200 {\it Kepler} stars spanning evolutionary stages from the main sequence to the
red-giant branch and found the radii to be accurate to within 5\%.
{\it Gaia} parallaxes have also been used to test the radii of the 66 main-sequence
stars of the {\it Kepler} LEGACY sample \citep{Lund2017,SilvaAguirre2017} which were
derived by fitting the individual frequencies to stellar models.
This test showed a systematic offset between seismic and {\it Gaia} parallaxes, with the
seismic parallaxes being about 0.25~mas larger on average.
This is in good agreement with the findings of \citet[hereafter ST16]{Stassun2016} who
derived parallaxes based on radii of eclipsing binaries.
\citet{Davies2017} compared {\it Gaia} parallaxes to those of red clump stars by adopting a
common absolute magnitude based on litterature values.
They found an offset which increases with parallax and reaches the
\citetalias{Stassun2016} value for the largest parallaxes of their sample
(${{\sim}1.6}$~mas).

It is possible that some of these offsets are caused partly by the temperature scale
which affects both the radii themselves and the transformation between radii and
parallaxes.
Indeed, \citet{Huber2017} showed that the use of different temperature scales for
main-sequence and subgiant stars can change the mean parallax difference by about 2\% for
results based on the scaling relations.
They also found that the offsets identified by \citetalias{Stassun2016} and
\citet{Davies2017} were overestimated for stars with parallaxes ${\lesssim 5}$~mas.

In this paper we take a closer look at the comparison between seismic and {\it Gaia}
parallaxes for the {\it Kepler} main-sequence and subgiant stars.
This includes an investigation of the impact of the temperature scale on the results of
both the scaling relations and the model fits to individual frequencies.
We rederive stellar parameters for the LEGACY sample using different effective
temperatures and extend the sample by including the Kages stars
\citep{Davies2016,SilvaAguirre2015}.
Additionally, the stellar parameters given by the scaling relations are compared to the
ones obtained from the analysis of individual frequencies.
Based on the assumption that the individual frequencies give the most accurate stellar
parameters obtainable by asteroseismology, we seek possible corrections to the scaling
relations, mainly as a function of effective temperature.

\section{Samples and Data}
\label{sec:samples}
We consider two different (but partially overlapping) samples of {\it Kepler} stars
which will be referred to as the main sample and the frequency sample.
The main sample is defined as all stars which have $\Delta\nu$ and
$\nu_{\mathrm{max}}$ from \citet{Chaplin2014}, effective temperatures and metallicities
obtained from spectroscopy by \citet{Buchhave2015}, and parallaxes from the GDR1
\citep{Lindegren2016}.
The frequency sample consists of all of the stars in the Kages and LEGACY samples with
parallaxes in the GDR1, except for the two Kages stars that showed signs of mixed dipole
modes.
For most of this sample, $T_{\mathrm{eff}}$ and $[\mathrm{Fe}/\mathrm{H}]$ are also taken
from \citet{Buchhave2015}; however, not all of the LEGACY stars were included in that
study.
For the ones missing (11 stars in total), $T_{\mathrm{eff}}$ and
$[\mathrm{Fe}/\mathrm{H}]$ are taken from the same spectroscopic sources as used in the
original LEGACY analysis by \citet{SilvaAguirre2017}.
This results in a main sample of 449 dwarf and subgiant stars with a completely
homogeneous set of observables, and a frequency sample of 86 dwarf stars.
Additionally, for all stars analysed, photometry in the infrared $JHK_{s}$ filters
was collected from the Two Micron All-Sky Survey (2MASS) catalogue \citep{Skrutskie2006},
and photometry in the $griz$ filters, as well as $E(B-V)$, were collected from the Kepler
Input Catalogue (KIC; \citealt{Brown2011}).
The $griz$ photometry was transformed to the SDSS scale using the corrections by
\citet{Pinsonneault2012a}.

The two samples have 52 stars in common for which the same temperatures and metallicities
are used, but the asteroseismic observables $\Delta\nu$ and
$\nu_{\mathrm{max}}$ have been determined separately.
For the main sample, the asteroseismic observables have been determined based on one
month of short cadence {\it Kepler} data for each star (see \cite{Chaplin2014} for
further details).
The stars of the frequency sample are a sub-sample of the stars which were chosen to be
observed for longer, and for most of them the asteroseismic observables are based on at
least 12 months of data.
As a result, the values from the frequency sample are  an order of magnitude more
precise.
The median relative uncertainties for the stars in common are 0.2\% in $\Delta\nu$ and
0.5\% in $\nu_{\mathrm{max}}$ for the frequency sample, and for the main sample the
corresponding values are 2.0\% and 4.3\%.
With 449 stars in the main sample, 86 in the frequency sample, and 52 in common between
them, there are 483 unique stars which will be referred to collectively as the full
sample.

\begin{figure}
  \center
  \includegraphics[width=\columnwidth]{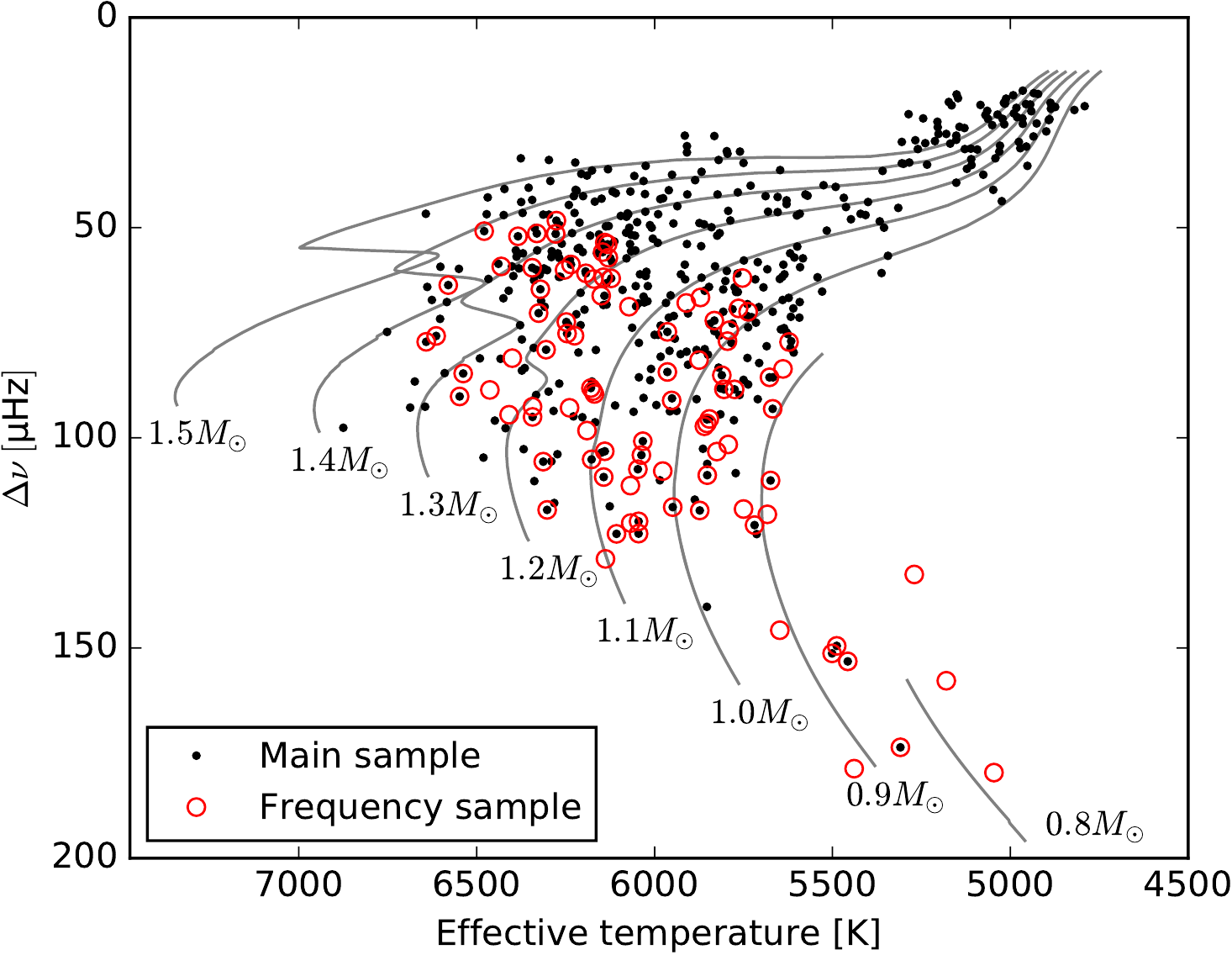}
  \caption{Observed $\Delta\nu$ and spectroscopic temperature of the main and frequency
           samples.
           The lines show evolutionary tracks computed with GARSTEC for models
           with solar metallicity and masses as labelled.}
  \label{fig:samples}
\end{figure}

Both samples are shown in \autoref{fig:samples} in the space of observed $\Delta\nu$ and
spectroscopic $T_{\mathrm{eff}}$ along with evolutionary tracks computed using the
Garching Stellar Evoluion Code (GARSTEC; \citealt{Weiss2008}).
With the visual aid of the evolutionary tracks, it is clear that the stars in the
frequency sample are generally less evolved than the stars in the main sample.
This is mainly due to the fact that the stars of the frequency sample have been selected
to be on the main sequence in order to avoid the complications of modelling mixed dipole
modes which show up in the oscillation spectra of more evolved stars.

\begin{figure}
  \center
  \includegraphics[width=\columnwidth]{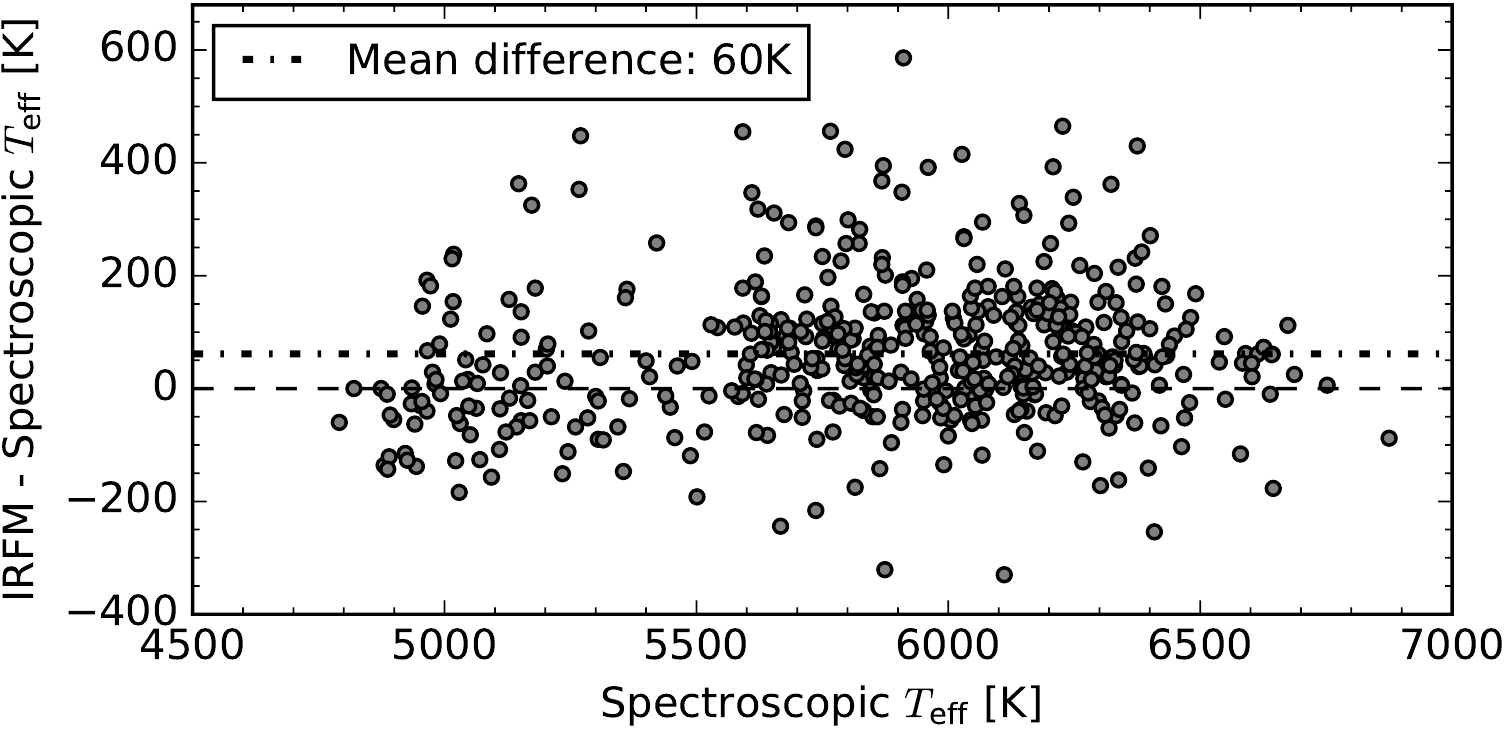}
  \caption{Difference between the two sets of temperatures (IRFM and spectroscopic) which
           have been collected for the full sample. On average, the IRFM temperatures
           are higher and the dot-dashed line shows the mean difference which is 60~K.}
  \label{fig:teff_comparison}
\end{figure}

Finally, a second effective temperature has been derived for the full sample using the
Infrared Flux Method (IRFM) as implemented by \cite{Casagrande2006,Casagrande2010}
(the procedure is described in \autoref{sec:distances}).
In \autoref{fig:teff_comparison} the differences between the IRFM and spectroscopic
temperatures are shown.
The scatter is quite large (about 150~K) and, on average, the IRFM temperatures are
higher than the spectroscopic ones by about 60~K;
however, this mean offset does not apply to the entire range of temperatures of the
sample.
If only the stars with temperatures below 5500~K are considered, which are mostly
subgiants (see \autoref{fig:samples}), the mean offset is instead around just 10~K
(but still with a large scatter).
For stars with temperatures above 5500~K, the mean offset agrees with the
overall mean of 60~K which is also the case for both the main and frequency
samples when considered individually.

\section{Methods}
\label{sec:methods}

\subsection{Radii}
\label{sec:radii}
The scaling relations for the average asteroseismic parameters
$\Delta\nu$ and $\nu_{\mathrm{max}}$ are given by \citep{Ulrich1986,Brown1991b}
\begin{align}
  \frac{\Delta\nu}{\Delta\nu_{\sun}} &\simeq \left(\frac{M}{M_{\sun}}\right)^{1/2}
                                              \left(\frac{R}{R_{\sun}}\right)^{-3/2}
                                              \label{eq:dnu_scal} \; ,
  \\
  \frac{\nu_{\mathrm{max}}}{\nu_{\mathrm{max},\sun}} &\simeq
                          \left(\frac{M}{M_{\sun}}\right)
                          \left(\frac{R}{R_{\sun}}\right)^{-2}
                          \left(\frac{T_{\mathrm{eff}}}{T_{\mathrm{eff},\sun}}\right)^{-1/2}
                          \label{eq:numax_scal} \; ,
\end{align}
which can be rearranged to give the scaling relations for the radius and mass
\begin{align}
  \frac{R}{R_{\sun}} &\simeq \left(\frac{\nu_{\mathrm{max}}}{\nu_{\mathrm{max},\sun}}\right)
                              \left(\frac{\Delta\nu}{\Delta\nu_{\sun}}\right)^{-2}
                              \left(\frac{T_{\mathrm{eff}}}{T_{\mathrm{eff},\sun}}\right)^{1/2}
                              \; , \label{eq:R_scal} \\
  \frac{M}{M_{\sun}} &\simeq \left(\frac{\nu_{\mathrm{max}}}{\nu_{\mathrm{max},\sun}}\right)^{3}
                              \left(\frac{\Delta\nu}{\Delta\nu_{\sun}}\right)^{-4}
                              \left(\frac{T_{\mathrm{eff}}}{T_{\mathrm{eff},\sun}}\right)^{3/2}
                              \; . \label{eq:M_scal}
\end{align}
The scaling relations have been applied to both stellar samples using the two different
effective temperature scales resulting in two sets of stellar parameters.
The solar values are taken to be
$\nu_{\mathrm{max,\sun}} = 3090~\mathrm{\upmu Hz}$,
$\Delta\nu_{\sun} = 135.1~\mathrm{\upmu Hz}$ \citep{Huber2011a}, and
$T_{\mathrm{eff,\sun}} = 5777~\mathrm{K}$.

In order to take advantage of the individual frequencies, the stars of the frequency
sample have been fitted to a grid of GARSTEC stellar models with theoretical oscillation
frequencies computed using the Aarhus adiabatic oscillation package (ADIPLS,
\citealt{Christensen-Dalsgaard2008}).
The stellar models have been computed with the OPAL equation of state \citep{Rogers2002},
OPAL opacitites \citep{Iglesias1996} with low-temperature opacities by
\citet{Ferguson2005}, the NACRE compilation of nuclear reaction rates \citep{Angulo1999},
and the solar mixture of \citet{Grevesse1998}.
Convection is implemented in the mixing length formalism and we use a solar-calibrated
mixing length parameter of $\alpha_{\mathrm{mlt}} = 1.7917$.
The solar-calibrated initial composition of $Y_{i} = 0.2705$ and $Z_{i} = 0.0189$ has
been combined with the Big Bang nucleosynthesis values of $Y_{0} = 0.2482$ and
$Z_{0} = 0$ \citep{Steigman2010a} to get a helium enrichment law of
$\Delta Y / \Delta Z = 1.179$.
This has been used in all models to define the initial helium abundance for a given
initial metallicity $[\mathrm{Fe}/\mathrm{H}]$.

Two grids of stellar models have been computed: one including microscopic
diffusion and the other including convective overshoot using the diffusion formalism
implemented in GARSTEC with an efficiency parameter of $f=0.016$
(see \citealt[section 3.1.5]{Weiss2008}).
The diffusion grid spans masses of 0.70--1.30~$M_{\sun}$ and the overshoot grid spans
masses of 1.00--1.80~$M_{\sun}$, both in steps of 0.01~$M_{\sun}$.
Initial metallicities from -0.65 to +0.55~dex in steps of 0.05~dex are included in the
overshoot grid and for the diffusion grid the initial values have been increased
slightly, thus covering the range -0.60 to +0.65~dex, in order to account for the
decrease in surface metallicity with evolution.
Both grids cover $\Delta\nu$, as calculated from \autoref{eq:dnu_scal}, in the range
13--180~$\upmu$Hz, which spans the entire observed range including some room for
inaccuracies in the scaling relations.
All stars were fitted to both model grids and for the ones that returned masses in the
range of overlap (1.00--1.30~$M_{\sun}$),
we choose, by visual inspection, the fit for which the probability distribution of
the mass is not cut off by the edge of the grid.
If none of the two fits hit the edge, we compare the observed and modelled oscillation
frequencies, effectively choosing the grid for which the best fitting model has the
highest likelihood.
Small individual diffusion grids were computed for the three stars with metallicities
lower than -0.60~dex.

For all fits to the grids, metallicities and temperatures are included in addition to the
asteroseismic observables.
Like for the scaling relations, this gives two sets of stellar parameters corresponding
to the use of the two different temperature scales.
Theoretical stellar oscillations are affected by systematic errors caused by improper
modelling of the outer stellar layers in current stellar models.
Therefore, when fitting the observed oscillation frequencies to the model grids, we have
to correct the model frequencies or use certain ratios of frequency differences which
have been shown to be insensitive to the surface layers \citep{Roxburgh2003}.
Using the Bayesian Stellar Algorithm (BASTA, see \citealt{SilvaAguirre2015}) we have
fitted the stars to both frequency ratios and to individual frequencies after applying
the correction introduced by \cite{Ball2014} for the surface effect.
These two methods give very similar results with a mean radius difference of less than
0.1\% and a scatter of 0.7\%.
Thus, the choice between fitting to frequency ratios or corrected individual frequencies
has no significant impact on the final results, and the stellar parameters based on fits
to frequency ratios will be used in the following.

The grid-based method has also been applied to the stars of the main sample by fitting
the observed $\Delta\nu$ to the mean frequency separation of the radial model
frequencies, $\Delta\nu_{\mathrm{fit}}$, calculated as described by \citet{White2011a}.
This quantity is also sensitive to the surface effect; therefore, we have scaled the grid
values by the ratio of the observed solar frequency separation,
$\Delta\nu_{\sun} = 135.1~\mathrm{\upmu Hz}$, and
that of the solar model calibration,
$\Delta\nu_{\mathrm{fit,sunmod}} = 136.1~\mathrm{\upmu Hz}$.
By applying this scaling, the surface effect is assumed to increase the value of
$\Delta\nu$ by a constant fraction of 1.007 for all stars.
Note that $\nu_{\mathrm{max}}$ has not been included in these model fits in order to get a
set of stellar parameters independent of the scaling relations.
It turns out, however, that it makes no difference whether it is included or not because
the typical relative uncertainty on $\nu_{\mathrm{max}}$ of 4.3\% for the stars of the
main sample is too large to impose any significant constraint on the models.

\subsection{Distances}
\label{sec:distances}
In order to calculate asteroseismic distances, and thereby parallaxes, we have used the
distance modulus based on the magnitudes in the $griz$ and $JHK_{s}$ photometric filters,
and the implementation is divided into two steps.
The first step of the calculation is to determine the reddening in a way to make it
consistent with the final distance.
This is done by an iterative procedure starting with the value of $E(B-V)$ from the KIC.
This value of the color excess is used as input for the bolometric correction software by
\citet{Casagrande2014} (as implemented in BASTA) together with the observed surface
properties $T_{\mathrm{eff}}$ (spectroscopic or IRFM depending on which was used to
determine the stellar properties), $[\mathrm{Fe}/\mathrm{H}]$, and the surface gravity
$\log g$ obtained from the asteroseismic analysis.
The output is the extinction-corrected bolometric corrections for each of the photometric
filters.
With each observed magnitude and theoretical bolometric correction, the distance is
calculated using the luminosity from asteroseismology and the absolute bolometric
magnitude of the Sun $M_{\mathrm{bol},\sun} = 4.75$.
The final distance is determined by taking the median of the values from the different
filters.
With this distance, the color excess is updated using the 3D dust map by
\citet{Green2015} and the process is iterated twice, at which point the color excess has
converged to within 0.01~mag.

The second step of the distance calculation is to determine the distances and
uncertainties using the final color excess obtained in step one.
For this purpose, we apply Monte Carlo sampling using the uncertainties on all input
parameters assuming Gaussian error distributions.
The values and uncertainties for $T_{\mathrm{eff}}$, $[\mathrm{Fe}/\mathrm{H}]$, and the
$JHK_{s}$ magnitudes are taken from the input catalogues.
No individual uncertainties are given for the $griz$ magnitudes in the KIC so they are
simply all set to 0.02~mag which is the level of precision found by \citet{Brown2011}
based on repeatability of the photometry.
Uncertainties on the luminosity and surface gravity are taken from the results of the
asteroseismic analysis.

We have also combined the asteroseismic scaling relations with the IRFM in the way
described by \citet{SilvaAguirre2012} to derive a set of self-consistent temperatures
(the ones introduced in \autoref{sec:samples}), radii, angular diameters, and hence
distances.
For this procedure the 3D dust map by \citet{Green2015} has also been used iteratively.
We applied the IRFM twice using different input photometry in the visible.
One set was obtained using the $B_{T}$ and $V_{T}$ bands of the Tycho-2 catalogue
\citep{Hog2000a} and the other set made use of the $griz$ bands from the KIC.
In both cases, the 2MASS $JHK_{s}$ bands were used in the infrared.
This was done since either set of visible photometry may be problematic in some cases.
For example, for the two brightest stars of this study, 16 Cygni A \& B, we find the
parallaxes to be underestimated when using the KIC photometry (compared to the {\it Gaia}
observations), but not when using Tycho-2 photometry, which indicates problems with the
photometry in the $griz$ filters for these stars.
On the other hand, the quality of the Tycho-2 photometry is worse for the faintest stars.
An initial comparison between the two sets of temperatures showed large discrepancies
with differences above 200~K for about 15\% of the stars.
Therefore, we decided to use a third set of temperatures for comparison in order to
choose, on a star-by-star basis, which of the two temperatures from the IRFM to use.
The third set was calculated from 2MASS photometry only, which is usually good over a
wide range of magnitudes, using a color-temperature calibration linking $J-K_{s}$ to
$T_{\mathrm{eff}}$.
For each star, the final temperature was chosen to be the IRFM temperature which was
closest to the one from the 2MASS color calibration.
With this criterion, the KIC ($griz$) set of temperatures and angular diameters was used
for 53\% of the stars and the Tycho-2 ($B_{T}V_{T}$) set was used for the remaining
47\%.
We could of course have used the 2MASS temperatures directly; however, this would defeat
the purpose of using the IRFM which is to obtain a set of self-consistent temperatures,
radii, and distances.

In principle, the IRFM parameters should be derived anew when using the individual
frequencies, instead of the scaling relations, to derive stellar parameters.
However, in practice, the radii which are used to get the distances do not change enough
for the changes in reddening to be significant.
A test showed that for the majority of the frequency sample the difference in $E(B-V)$ is
at the level of 0.001~mag.
This level of agreement translates to a difference in $T_{\mathrm{eff}}$ of just a few
kelvin which is well below the statistical uncertainties.

\section{Results}
\label{sec:results}

\subsection{Parallax comparisons for the frequency sample}
\label{sec:par_freqs}
\begin{figure*}
  \center
  \includegraphics[width=\textwidth]{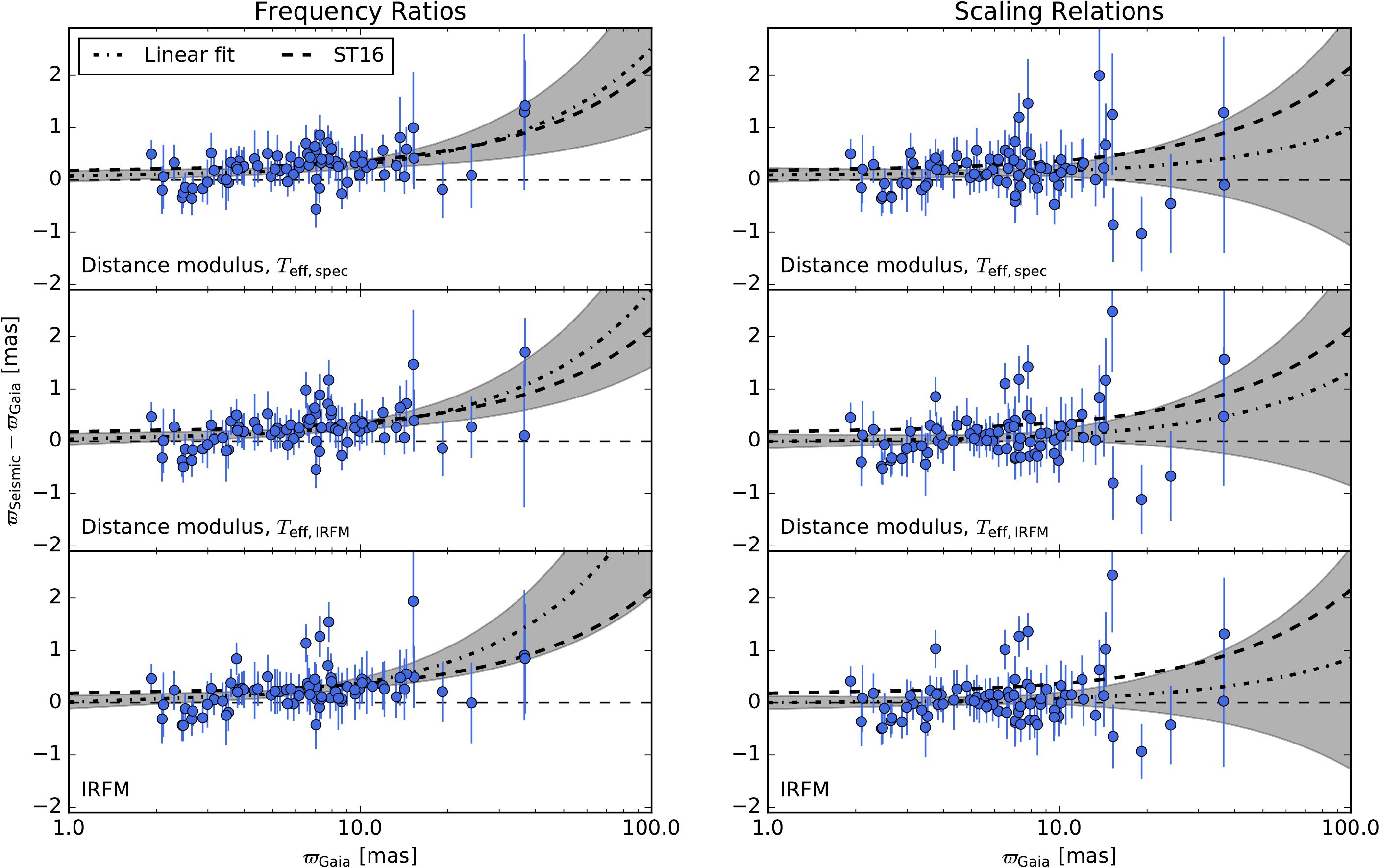}
  \caption{Absolute differences between seismic and {\it Gaia} parallaxes as a function of {\it Gaia}
           parallax for the frequency sample. The dot-dashed lines are linear fits to the
           data which are weighted according to the uncertainties and the thick dashed
           line is the linear relation found by
           \citetalias{Stassun2016} based on radii of eclipsing binaries (the lines curve
           due to the logarithmic x-axis). Grey shaded areas indicate the
           1$\sigma$-region of the linear fits. The left- and right-hand columns show
           results based on frequency ratio fits and the scaling relations, respectively.
           The rows show different methods for calculating seismic distances (and hence
           parallaxes) as labelled.}
  \label{fig:par_diff_freqs}
\end{figure*}
Based on the seismic distances, $d_{\mathrm{seis}}$, we simply calculate the seismic
parallaxes as $\varpi_{\mathrm{seis}} = 1/d_{\mathrm{seis}}$.
In \autoref{fig:par_diff_freqs} the absolute differences between the seismic and {\it Gaia}
parallaxes are plotted as a function of the {\it Gaia} parallaxes for the frequency sample.
The columns show the two different methods used to derive stellar properties, and the
rows show the different methods for calculating distances.
A number of outliers were identified in the original study of the LEGACY sample, and all
of the stars they found to be problematic have been excluded from this comparison.
They quoted several sources to argue that the offsets are mainly due to contaminated
photometry.
For example, some of the stars are binaries and their flux measurements include the
contributions from both components (see the discussion in
\citealt[section 5.3]{SilvaAguirre2017}).

If we first consider the results based on fits to frequency ratios (left-hand column),
there is good agreement overall and individual differences are generally within the
statistical uncertainties.
However, there is also a trend towards seismic parallaxes being too large on average. 
For the results based on spectroscopic temperatures, the weighted mean difference is
$\Delta\varpi=(0.21\pm0.04)~\mathrm{mas}$, and the offset also seems to be distance
dependent as indicated by the linear fit to the data which is weighted according to the
plotted uncertainties.
Based on this fit, the difference as a function of parallax is approximately given by
\begin{equation} \label{eq:freqs_linfit_offset}
  \Delta\varpi = (0.03 \varpi + 0.03)~\mathrm{mas} \, .
\end{equation}
This is similar to what was found by \citetalias{Stassun2016} based on radii of eclipsing
binaries (dashed lines in the figure), but their linear relation predicts a slightly
larger offset at low parallaxes (with an intersection at 0.16~mas) which is just outside
the $1\sigma$-region of what is found here.
For parallaxes ${\gtrsim}10~\mathrm{mas}$ the two linear relations agree.
A distance dependent absolute offset means that the relative offset is constant across
all distances.
The weighted mean parallax ratio is $1.029\pm0.006$, meaning that the seismic parallaxes
are overestimated by about 3\% assuming that the {\it Gaia} parallaxes are unbiased.

An overestimation of the seismic parallaxes is exactly what we would expect to get if the
temperatures are underestimated.
Higher temperatures will lead to higher luminosities which increase the derived
distances and lowers the parallaxes.
Comparing the stellar parameters obtained with the two sets of temperatures, the increase
of about 60~K on average leads to best-fitting models which are ${\sim}2\%$ more
luminous.
The mean change in radius, however, is below 0.1\% which means that the entire increase
in luminosity is due to the best-fitting models being hotter.
Comparing the upper two panels, which differ only in the adopted temperatures, we see
that the parallaxes are still too high on average.
Surprisingly, with the IRFM temperatures, the weighted mean ratio has not decreased.
In fact, it has increased slightly (but not significantly) to $1.031\pm0.006$ for
the results using the distance modulus.
As argued above, higher luminosities should lead to lower parallaxes and better overall
agreement with the {\it Gaia} data.
Based on the proportionality from the distance modulus
$\varpi \propto (L/L_{\odot})^{-1/2}$, the parallaxes are expected to decrease by about
$1\%$ due to the luminosity increase of $2\%$.
If we instead compare the median ratios, the results are $1.041$ and $1.033$ with the
spectroscopic and IRFM temperatures, respectively.
This is more in line with our expectations and indicates that the parallaxes have
decreased overall, but the stars with high weights (i.e.\ low statistical uncertainties)
have retained their seismic parallax value.
In any case, there is a systematic offset at the level of $3\%$ regardless of the
adopted temperature scale.
A comparison between the two lower panels shows that the two different methods used to
calculate distances agree very well.
When comparing the two sets of seismic parallaxes directly, they are found to agree
within 0.2\% on average even though different photometry was used in the visible for
two-thirds of the stars (which is the fraction of the frequency sample for which the
Tycho-2 photometry was used to derive the IRFM temperatures).

\begin{figure*}
  \center
  \includegraphics[width=\textwidth]{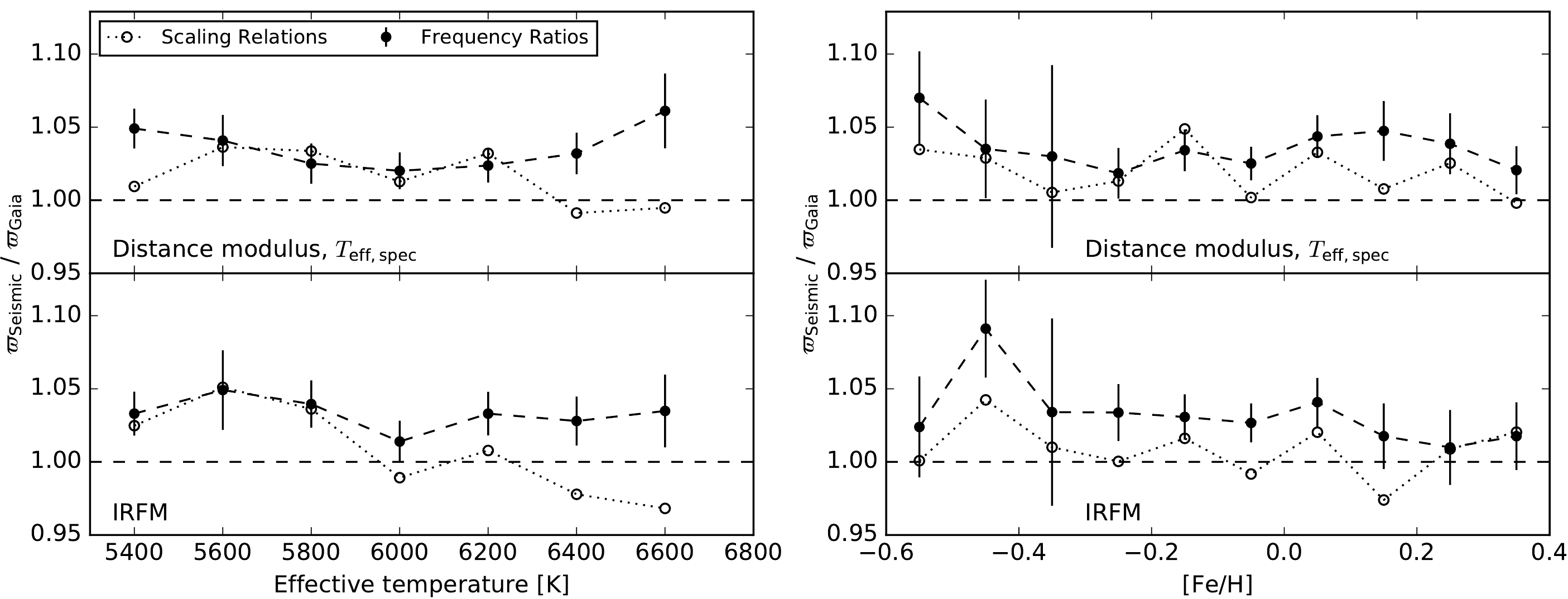}
  \caption{Ratios of seismic and {\it Gaia} parallaxes as a function of effective temperature
           (left) and metallicity (right) for the frequency sample. The data are shown as
           weighted mean bins with equal separation. Errorbars on the binned data are
           shown for the results based on frequency ratios to give a sense of the
           uncertainty at a given value on the x-axis. The most metal-poor star of the
           sample ($[\mathrm{Fe}/\mathrm{H}] = -0.92$) has been excluded in the
           right-hand panel.}
  \label{fig:par_diff_freqs_teff}
\end{figure*}

Turning to the results based on the scaling relations (right-hand column of
\autoref{fig:par_diff_freqs}), the mean parallax offset has decreased compared to the
results based on ratio fits.
For the spectroscopic temperatures, the weighted mean difference is
$\Delta\varpi=(0.14\pm0.05)~\mathrm{mas}$ which still implies that the seismic parallaxes
are too high.
In this case, however, the temperature scale does make a difference since it enters
explicitly into the scaling relations, and with IRFM temperatures and angular diameters
the offset decreases to just $\Delta\varpi=(0.05\pm0.05)~\mathrm{mas}$, consistent with
no offset.
Note also that for both temperature scales the slope of the linear fit is not
significantly different from zero, and for the IRFM temperatures the offset is clearly not
compatible with the results of \citetalias{Stassun2016} at low parallaxes.
This stands in contrast to the results based on ratio fits where the slope is significant
regardless of the adopted temperatures.

These results corroborate the findings of \citet{Huber2017}, namely that the hotter IRFM
temperatures lead to better agreement between seismic (from scaling relations) and {\it Gaia}
parallaxes, and that the offset identified by \citetalias{Stassun2016} may be
overestimated for low parallaxes.
However, the offset found by \citet{SilvaAguirre2017} for the LEGACY sample has not been
reduced by the inclusion of the Kages stars or by the change in temperature.
This leads to an interesting tension between the scaling relations giving good agreement
with the {\it Gaia} parallaxes and the fits to frequency ratios showing a constant offset of a
few percent.

The differences between the results from the scaling relations and the ratio fits, and
the impact of the temperature scale, become clearer when the results are plotted together
as a function of temperature as shown in the left-hand panel of
\autoref{fig:par_diff_freqs_teff}.
To avoid clutter, only the weighted mean bins are shown, and the error bars, which are of
similar magnitude for all data points at a given temperature, are only shown for the
ratio fits.
At the lowest temperatures, the scaling relations generally agree well with the ratio
fits for both temperature scales.
However, at higher temperatures the scaling relations deviate from the ratio fits and
result in lower parallaxes.
So the main reason that the parallaxes from the scaling relations show overall better
agreement with {\it Gaia} seems to be a deviation from the ratio fits at high temperatures.
There are also differences between the two applications of the scaling relation (upper
and lower panel) due to the different temperatures.
At high temperatures ($T_{\mathrm{eff}} > 5800~\mathrm{K}$) the IRFM results give lower
seismic parallaxes than the spectroscopic results and vice versa at the lower
temperatures.
This is what makes the IRFM results compatible with the {\it Gaia} data (on average), and it is
simply a reflection of differences in the temperature scales.

The right-hand panel of \autoref{fig:par_diff_freqs_teff} shows the parallax comparison
as a function of metallicity where no significant trends are found.
Here the difference between the scaling relations and the ratio fits show up as a
constant offset which does not depend on the metallicity.

Since the scaling relations are anchored to the solar values, they are naturally thought
to be most accurate for stars like the Sun.
Therefore, it is interesting that the parallaxes based on the scaling relations agree so
well with the ones from frequency ratios at near-solar temperatures and only deviate at
the highest temperatures considered in this study.
This indicates that either the asteroseismic radii are underestimated at temperatures
around that of the Sun, and the scaling relations become more accurate at higher
temperatures, or, which seems more likely, that there is a bias in either the {\it Gaia}
parallaxes or the calculation of seismic parallaxes, and that the scaling relations
overestimate the radii at high temperatures.
If the deviation between the two methods at high temperatures is indeed due to
inaccuracies in the scaling relations, we are still left with the problem of explaining a
3\% offset in seismic parallaxes based on fits to frequency ratios.

\begin{figure*}
  \center
  \includegraphics[width=\textwidth]{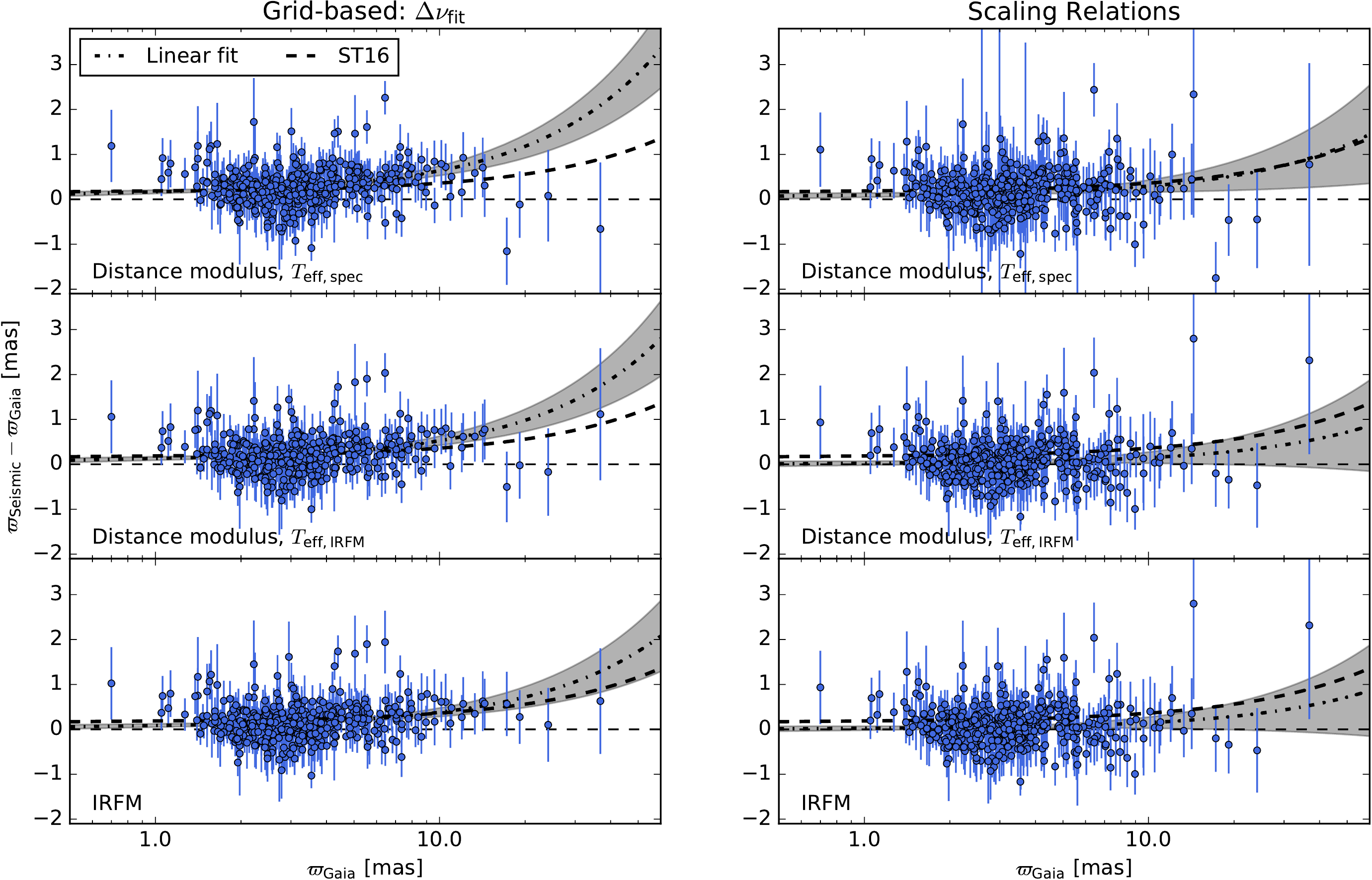}
  \caption{Same as \autoref{fig:par_diff_freqs} but for the main sample with the
           grid-based results based on $\Delta\nu_{\mathrm{fit}}$ instead of frequency
           ratios.}
  \label{fig:par_diff_scaling}
\end{figure*}

The {\it Gaia} parallaxes are known to suffer from position and colour dependent systematic
errors at a level of $\pm$0.3~mas \citep{Lindegren2016}.
This may explain part of the observed offset; however, it seems unlikely to explain a
constant fractional offset such as the one found here.
A fractional offset may instead be introduced in the conversion between seismic radii and
parallaxes.
\citet{Huber2017} tested the use of different ways to determine bolometric corrections
and found that it can lead to systematic errors of around 1\% in parallax due to the use
of different model atmospheres and methods for extracting the synthetic fluxes.
It is also possible that the extinction values of the 3D map are biased.
In order to decrease the seismic parallaxes and bring them into agreement with the {\it Gaia}
values, the degree of extinction has to be decreased.
However, the stars considered here are not very distant ($d < 1~\mathrm{kpc}$), and
the median extinction in the visible is only about 0.03~mag based on the
extinction map.
For comparison, a difference of about 0.06~magnitudes is necessary in order to explain
a relative difference of 3\% in parallax.

It is also worth discussing what could lead to underestimated seismic radii from the
ratio fits, if that is in fact the cause of the offset.
The most obvious sources of bias in the stellar parameters are the model physics that
have been fixed based on e.g.\ a solar calibration.
For example, the mixing length parameter $\alpha_{\text{mlt}}$ is fixed at the
solar-calibrated value of $1.791$ in all models.
However, 3D simulations of convection have shown that the convective efficiency varies
with temperature and surface gravity (e.g. \citealt{Trampedach2014}).
Additionally, there is a known degeneracy between the initial helium abundance and the
stellar mass \citep{SilvaAguirre2015} and, for the grids used in this study, the
initial helium abundance is fixed for any given metallicity by the enrichment law
$\Delta Y / \Delta Z = 1.179$.
Finally, the overshoot efficiency directly influences the obtained masses for stars with
convective cores.

Now the question is whether the effects of fixing the model physics are large enough to
explain the observed parallax offset.
In the original study of the LEGACY sample \citep{SilvaAguirre2017}, a number of different
stellar model grids and fitting algorithms were applied.
Most notably, a number of the algorithms left the values of $\alpha_{\text{mlt}}$ and
the initial helium abundance as free parameters to be optimized during the fit.
Even with this variety of methods, the overall agreement between the radii was good (with
typical mean offsets between the methods below 1\%) and the parallax offset compared to
the {\it Gaia} values was seen with every method.
Therefore, model physics are only able to explain part of the offset unless more
fundamental features of the models like the opacities or the equation of state need
adjustments.

All things considered, it is not possible to point to any one part of the analysis as the
sole source of the observed offset.
It may be caused by any combination of the different factors that have been discussed,
including of course the seismic radii.
What can be said, however, is that there are no indications of deviations from the
scaling relation for the radius by more than 5\% based on these parallax comparisons.

\subsection{Parallax comparisons for the main sample}
\label{sec:par_main}
In \autoref{fig:par_diff_scaling} the absolute differences between the seismic and {\it Gaia}
parallaxes are shown for the main sample.
One feature which must be mentioned is the tendency for the offset to be systematically
positive at the lowest parallaxes.
This happens because the {\it Gaia} parallaxes are less precise than the seismic ones, and
therefore scatter to lower values causing a positive offset.
Similarly, when they scatter to higher values the offset is negative, and the combined
effect is a diagonal edge at low parallaxes.

Like we saw for the ratio fits of the frequency sample, the grid-based method results in
absolute offsets which increase with parallax.
Due to the diagonal edge at low parallaxes, only stars with
$\varpi_{\mathrm{Gaia}} > 2.5~\mathrm{mas}$ have been included in the linear fits.
With spectroscopic temperatures the weighted mean difference is
$\Delta\varpi=(0.28\pm0.02)~\mathrm{mas}$, and the linear fit is given approximately by
\begin{equation}
  \Delta\varpi = 0.06 \varpi + 0.07 \, .
\end{equation}
This slope is double the value found for the ratio fits and by \citetalias{Stassun2016}.
The weighted mean relative offset has also doubled to $(6.0\pm0.4)\%$.

\begin{figure*}
  \center
  \includegraphics[width=\textwidth]{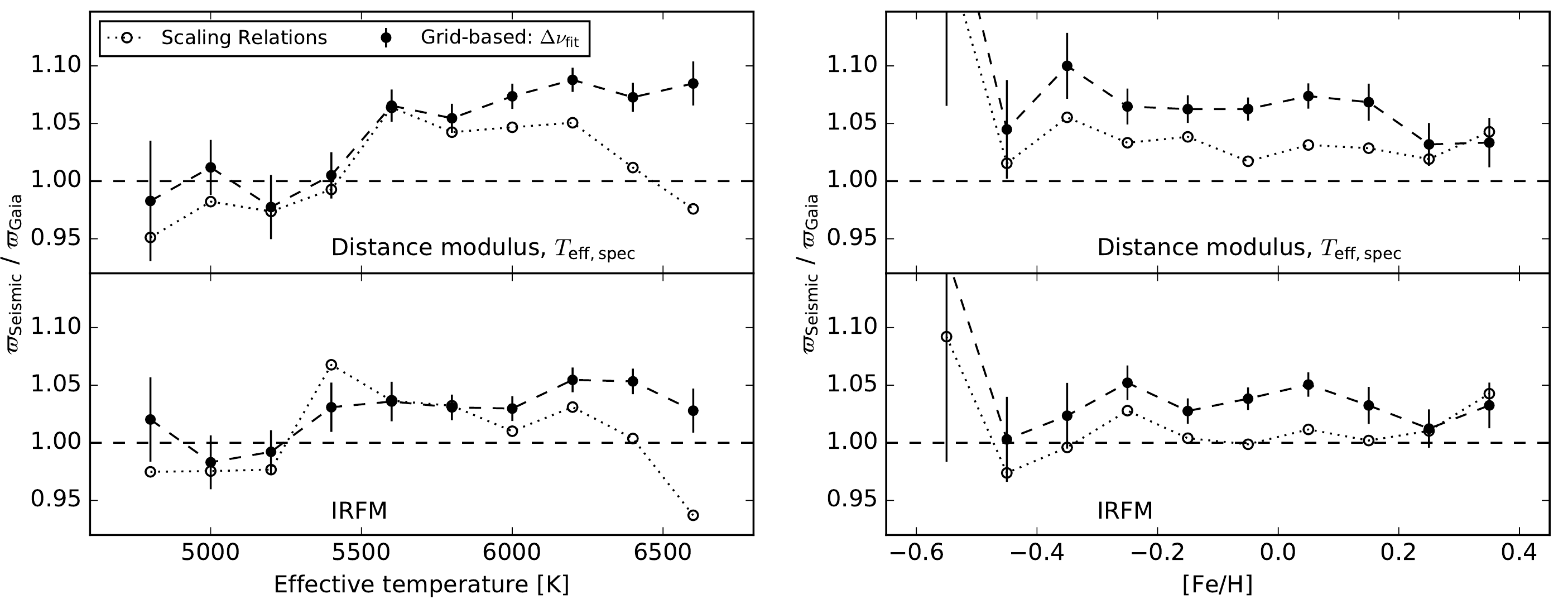}
  \caption{Same as \autoref{fig:par_diff_freqs_teff} but for the main sample with the
           grid-based results based on $\Delta\nu_{\mathrm{fit}}$ instead of frequency
           ratios. The two most metal-poor stars of the sample
           ($[\mathrm{Fe}/\mathrm{H}] = -0.92, -1.75$) have been excluded in the
           right-hand panel.}
  \label{fig:par_diff_scaling_teff}
\end{figure*}

The temperature carries more weight in the fits to $\Delta\nu_{\mathrm{fit}}$ of the main
sample than it did for the fits to frequency ratios, and the change to IRFM temperatures
has a larger impact.
Changing from spectroscopic temperatures and the distance modulus to IRFM temperatures
and angular diameters, the weighted mean parallax offset is reduced to $(3.3\pm0.4)\%$,
bringing it much closer to the results of the frequency sample and
\citetalias{Stassun2016}.
The combination of IRFM temperatures and the distance modulus gives an offset of
$(4.9\pm0.4)\%$ which falls right in between the two extremes.
Thus, the mean offset gradually decreases when going from the upper to the lower panels
of \autoref{fig:par_diff_scaling}.

Like for the frequency sample, the scaling relations lower both the overall offset
between the parallaxes and the slopes of the linear fits.
For the spectroscopic and IRFM temperatures, the weighted mean differences are
$\Delta\varpi = (0.16\pm0.02)~\mathrm{mas}$ and $(0.07\pm0.02)~\mathrm{mas}$,
respectively, in agreement with the offsets found with the scaling relations for the
frequency sample, but with lower uncertainties owing to the larger sample size.
The slope is only just significant within one standard deviation for the spectroscopic
temperatures, but for the IRFM temperatures it is insignificant.
It is also seen once again that the relation from \citetalias{Stassun2016} overestimates
the offset at low parallaxes compared to the results found here with both temperature
scales.

In \autoref{fig:par_diff_scaling_teff} the weighted mean bins from the scaling relations
and the fits to $\Delta\nu_{\mathrm{fit}}$ are compared directly for the two sets of
temperatures.
The scaling relations follow the fits to $\Delta\nu_{\mathrm{fit}}$ at low temperatures
where the seismic parallaxes show the best agreement with {\it Gaia}.
For higher temperatures, i.e.\ $T_{\mathrm{eff}}\gtrsim 6000~\mathrm{K}$ the two methods
begin to increasingly deviate.
Note how similar the trend is to the one for the frequency sample
(\autoref{fig:par_diff_freqs_teff}) if the temperatures below 5500~K are ignored.
In this region, both samples agree that the seismic parallaxes based on the scaling
relations change with temperature while the methods independent of the scaling relations
show a nearly constant offset.
Also like for the frequency sample, there are no significant trends with metallicity.

\subsection{Corrections to the scaling relations}
\label{sec:scal_corr}
If we assume that the parallax differences seen in \autoref{fig:par_diff_freqs_teff}
and \autoref{fig:par_diff_scaling_teff} are solely due to errors in the seismic radii,
the scaling relation for the radius needs a correction, depending on the temperature,
with a maximum value of about 5\%.
For example, using the results from the IRFM, the radii are, on average, underestimated
at the solar temperature and overestimated at the highest temperatures in the sample.
The best agreement between the parallaxes is found for the subgiants in the main sample
which suggests that little or no correction is needed to the scaling relation for these
stars.

The fact that both the parallaxes from the scaling relations and from frequency modelling
show an offset at the solar temperature (where the scaling relations are thought to be
accurate for dwarf stars) leads us to consider the alternate assumption that there is a
bias in either the Gaia parallaxes or in the quantities involved in transforming the
seismic radii into distances (i.e. bolometric corrections and extinctions).
The parallax comparisons show that the radii of the scaling relations deviate from those
of the grid-based analyses, which are independent of the scaling relations, at high
effective temperatures.
Even though the scaling relations give parallaxes which agree better with the {\it Gaia}
data on average, it seems most likely that the trend with temperature reflects a problem
with the scaling relations.
In the following, we attempt to quantify this temperature trend and correct the
scaling relations for it.
We will use the stellar parameters obtained from the ratio fits (with IRFM temperatures)
as reference values.
Despite the constant parallax offset of 3\%, these radii and masses are very precise and
they are internally consistent within the adopted stellar models.
Thus we consider them to be the best available substitute for the true values.
However, it should be kept in mind that the validity of the corrections we obtain
for the scaling relations depend on the assumption that the parallax offset of 3\%,
between parallaxes from {\it Gaia} and from fitting to frequency ratios, is not due to
systematic errors in the seismic radii.
We also assume that the temperature trend is not due to a potential temperature dependent
bias in the IRFM temperatures.

We take the stellar radii, masses, and temperatures from the best-fitting models and
calculate the model values $\Delta\nu_{\mathrm{scaling}}$ and
$\nu_{\mathrm{max,scaling}}$ based on the scaling relations (\autoref{eq:dnu_scal} and
\autoref{eq:numax_scal}).
It is interesting whether or not the deviations between $\Delta\nu_{\mathrm{scaling}}$,
$\nu_{\mathrm{max,scaling}}$ and the observed values depend on the physical parameters of
the stars.
If they do, it is possible to define corrections to the scaling relations which will
bring the radii and masses calculated from them into line with the ones found from the
ratio fits.
Since this analysis does not depend on the {\it Gaia} parallaxes we have included the LEGACY
and Kages stars which were not a part of the GDR1, increasing the sample size from 86 to
95 stars.

As we saw in the parallax comparisons, the difference between the radii obtained from the
ratio fits and those given by the scaling relations varies as a function of temperature.
This suggests that corrections to the scaling relations for $\Delta\nu$ and
$\nu_{\mathrm{max}}$ should at least depend on temperature.
For $\Delta\nu$ this is no surprise since previous studies (e.g. \citealt{White2011a})
already showed that the difference between $\Delta\nu_{\mathrm{fit}}$ and
$\Delta\nu$ from the scaling relation is mainly a function of temperature.
However, using the masses and radii of the ratio fits it will be possible to investigate
potential corrections to the scaling relation for $\nu_{\mathrm{max}}$ as well.
Defining the two functions $f_{\Delta\nu}(T_{\mathrm{eff}})$ and
$f_{\nu_{\mathrm{max}}}(T_{\mathrm{eff}})$ as the temperature dependent corrections to
the scaling relations for $\Delta\nu$ and $\nu_{\mathrm{max}}$, respectively, the
corrected versions of Equations~\ref{eq:dnu_scal} and \ref{eq:numax_scal} are given by
\begin{align}
  \frac{\Delta\nu}{\Delta\nu_{\sun}} &\simeq \left(\frac{M}{M_{\sun}}\right)^{1/2}
                                              \left(\frac{R}{R_{\sun}}\right)^{-3/2}
                                              f_{\Delta\nu}(T_{\mathrm{eff}})
                                              \label{eq:dnu_scal_corr} \; ,
  \\
  \frac{\nu_{\mathrm{max}}}{\nu_{\mathrm{max},\sun}} &\simeq
                          \left(\frac{M}{M_{\sun}}\right)
                          \left(\frac{R}{R_{\sun}}\right)^{-2}
                          \left(\frac{T_{\mathrm{eff}}}{T_{\mathrm{eff},\sun}}\right)^{-1/2}
                          f_{\nu_{\mathrm{max}}}(T_{\mathrm{eff}})
                          \label{eq:numax_scal_corr} \; ,
\end{align}
from which the corrected scaling relations for the mass and radius follow
\begin{align}
  \frac{M}{M_{\sun}} &\simeq \left(\frac{\nu_{\mathrm{max}}}{\nu_{\mathrm{max},\sun}}\right)^{3}
                              \left(\frac{\Delta\nu}{\Delta\nu_{\sun}}\right)^{-4}
                              \left(\frac{T_{\mathrm{eff}}}{T_{\mathrm{eff},\sun}}\right)^{3/2}
                              f_{\nu_{\mathrm{max}}}^{-3}(T_{\mathrm{eff}})
                              f_{\Delta\nu}^{4}(T_{\mathrm{eff}})
                              \; , \label{eq:M_scal_corr} \\
  \frac{R}{R_{\sun}} &\simeq \left(\frac{\nu_{\mathrm{max}}}{\nu_{\mathrm{max},\sun}}\right)
                              \left(\frac{\Delta\nu}{\Delta\nu_{\sun}}\right)^{-2}
                              \left(\frac{T_{\mathrm{eff}}}{T_{\mathrm{eff},\sun}}\right)^{1/2}
                              f_{\nu_{\mathrm{max}}}^{-1}(T_{\mathrm{eff}})
                              f_{\Delta\nu}^{2}(T_{\mathrm{eff}})
                              \; . \label{eq:R_scal_corr}
\end{align}
The combinations of correction factors in these equations can be defined as correction
factors for the mass $f_M = f_{\nu_{\mathrm{max}}}^{-3} f_{\Delta\nu}^{4}$, and the
radius $f_R = f_{\nu_{\mathrm{max}}}^{-1} f_{\Delta\nu}^{2}$.
Corrections that apply to the frequency sample can be obtained by comparing the observed
and modelled asteroseismic parameters as a function of temperature.
Only the stars with effective temperatures higher than 5500~K will be used
in this analysis due to the poor sample size at lower temperatures.

\begin{figure}
  \center
  \includegraphics[width=\columnwidth]{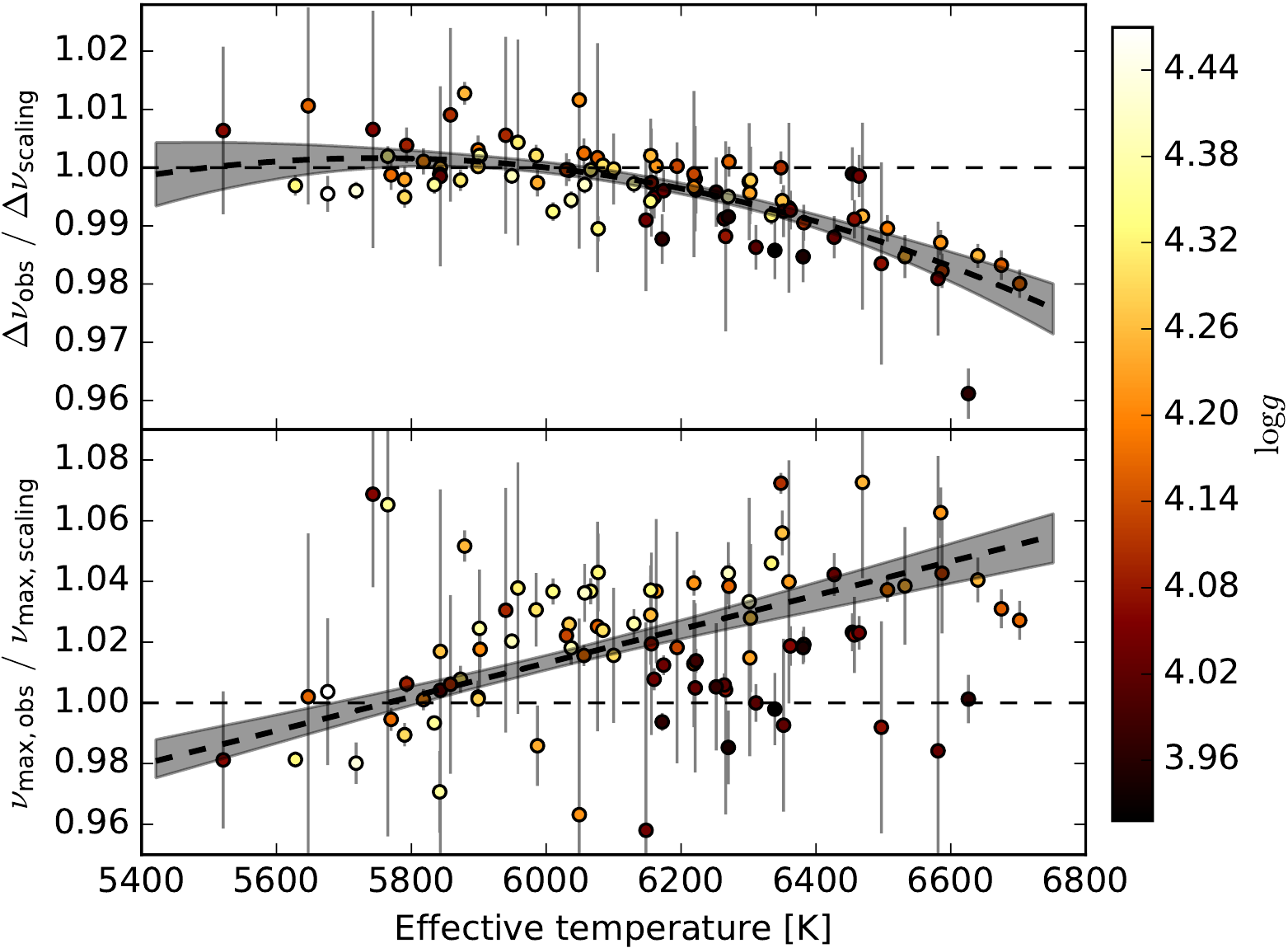}
  \caption{Ratio of observed and scaling relation $\Delta\nu$ (upper panel) and
           $\nu_{\mathrm{max}}$ (lower panel) as a function of IRFM effective temperature
           with error bars indicating the uncertainties on the observed values. The
           scaling relation values are taken from the best-fitting models of the ratio
           fits and depend on the models' masses, radii, and temperatures. The data are
           color coded according to surface gravity and the dashed lines are the
           best-fitting relations discussed in the text. The shaded areas indicate the
           bootstrapped uncertainties of the fits.}
  \label{fig:scaling_dnu_numax_corr}
\end{figure}

In \autoref{fig:scaling_dnu_numax_corr}, the ratios of observed and scaling relation
$\Delta\nu$ and $\nu_{\mathrm{max}}$ are shown as a function of observed effective
temperature and color coded by surface gravity.
The error bars only include the statistical uncertainty on the observed values.
Inspired by the functional form of the correction found by \citet{White2011a} a second
order polynomial has been fitted to the data for $\Delta\nu$.
The fit follows the data nicely, and the residual scatter is at a level of ${\sim}1\%$.
For this limited range of surface gravities, there is no strong indication that the
offset depends on the value of $\log g$.
Based on the fit, the correction factor for $\Delta\nu$ is given by
\begin{equation} \label{eq:dnu_corr_freqs}
  f_{\Delta\nu}(T_{\mathrm{eff}}) =
             - 2.52 \left(\frac{T_{\mathrm{eff}}}{10^4\mathrm{K}}\right)^{2}
             + 2.90 \left(\frac{T_{\mathrm{eff}}}{10^4\mathrm{K}}\right)
             + 0.17 \; ,
\end{equation}
which applies in the range of temperatures and surface gravities covered in the figure.
This expression gives a small correction of
$f_{\Delta\nu}(T_{\mathrm{eff},\sun}) = 1.002$ at the solar temperature which is not
ideal considering that the scaling relations are anchored to the solar values.
One could add a constraint on the fit to make it hit unity at the solar temperature, but
since the deviation is within the uncertainty (shaded area in the figure), we have chosen
not to do so.

\begin{figure}
  \center
  \includegraphics[width=\columnwidth]{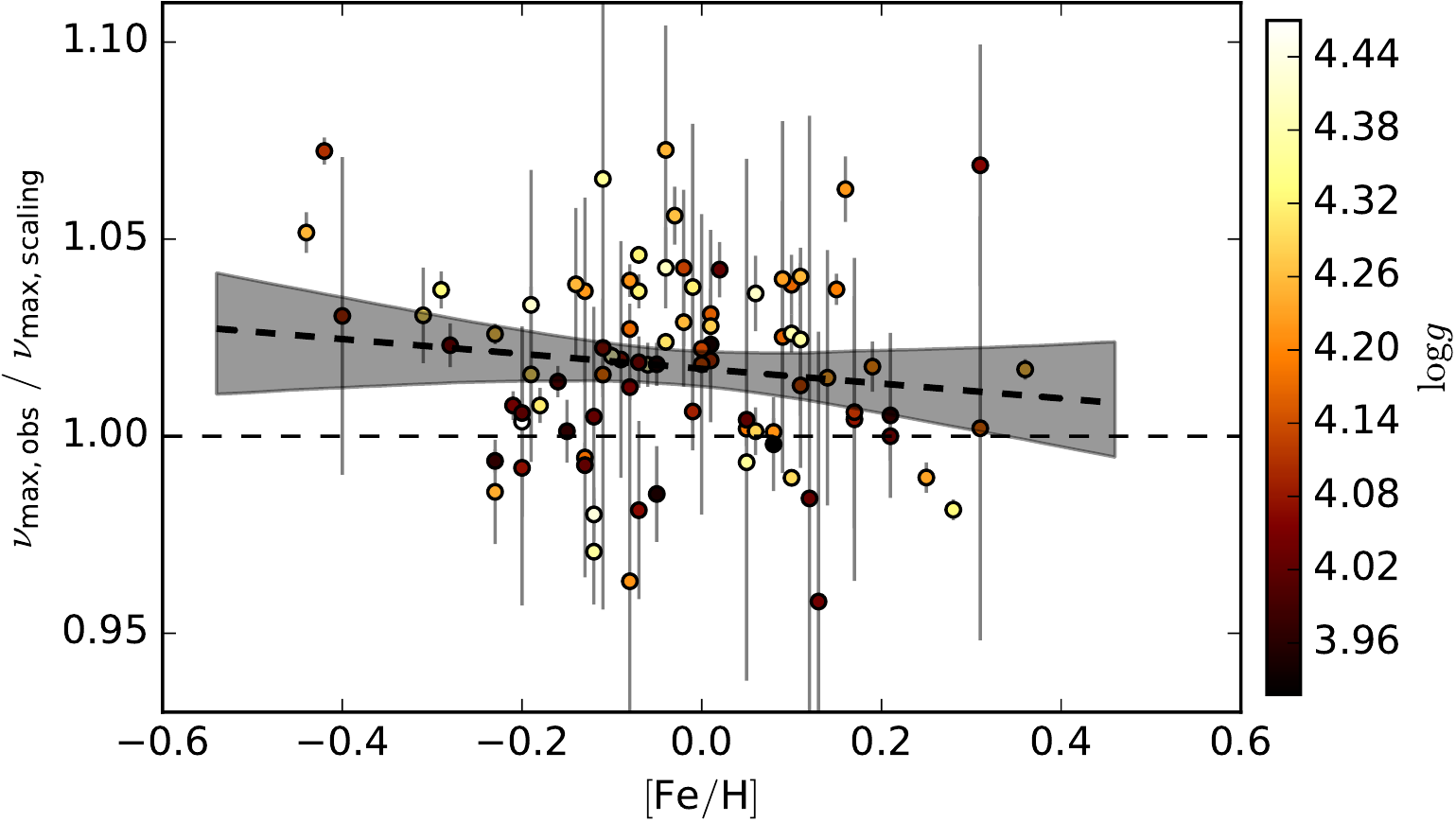}
  \caption{Ratio of observed and scaling relation $\nu_{\mathrm{max}}$ as a function of
           spectroscopic metallicity with errorbars indicating the uncertainty on the
           observed values. The scaling relation values are taken from the best-fitting
           models of the ratio-fits and depend on the models' masses, radii, and
           temperatures (using the IRFM set). The data are color coded according to
           surface gravity and the dashed line is the best linear fit. The shaded area
           indicates the bootstrapped uncertainty of the fit.}
  \label{fig:numax_io_feh}
\end{figure}

For $\nu_{\mathrm{max}}$ the uncertainties are larger and the data points are much more
scattered at any given temperature.
Although the correlation with temperature is less clear, the offset does seem to increase
slightly with increasing temperature.
The linear fit shown in the figure implies a correction factor of
\begin{equation} \label{eq:numax_corr_freqs}
  f_{\nu_{\mathrm{max}}}(T_{\mathrm{eff}}) =
               0.397 \left(\frac{T_{\mathrm{eff}}}{10^4\mathrm{K}}\right)
             + 0.771 \; .
\end{equation}
Like for the $\Delta\nu$ correction, the fit hits unity at the solar temperature within
the uncertainty and has not been constrained.
The large scatter around this relation is partly due to the uncertainties on the observed
values, but it may also suggest that the offset depends on more than just temperature.
Most of the stars with low surface gravities seem to fall below the linear fit which
indicates that the correlation can be improved by including a $\log g$ term.
A dependence on surface gravity is reasonable considering it is the other factor, besides
the temperature, which enters into the $\nu_{\mathrm{max}}$ scaling relation.
Additionally, a dependence on metallicity is not unthinkable considering that
\citet{Viani2017} recently pointed out that the approximation usually adopted for the
acoustic cut-off frequency $\nu_{\mathrm{ac}}$ (and therefore also $\nu_{\mathrm{max}}$)
lacks a factor related to the mean molecular weight.
However, no significant correlations with these parameters have been found.
A direct comparison between $\nu_{\mathrm{max,obs}} / \nu_{\mathrm{max,scaling}}$ and the
metallicity, showing no significant correlation, is given in \autoref{fig:numax_io_feh}.
Similarly, no significant correlation is found with $\log g$.
Thus, the dependence of the correction on surface gravity and metallicity is not very
strong, and the simple relation in \autoref{eq:numax_corr_freqs} has been adopted as the
correction to $\nu_{\mathrm{max}}$ in the following.

\begin{figure}
  \center
  \includegraphics[width=\columnwidth]{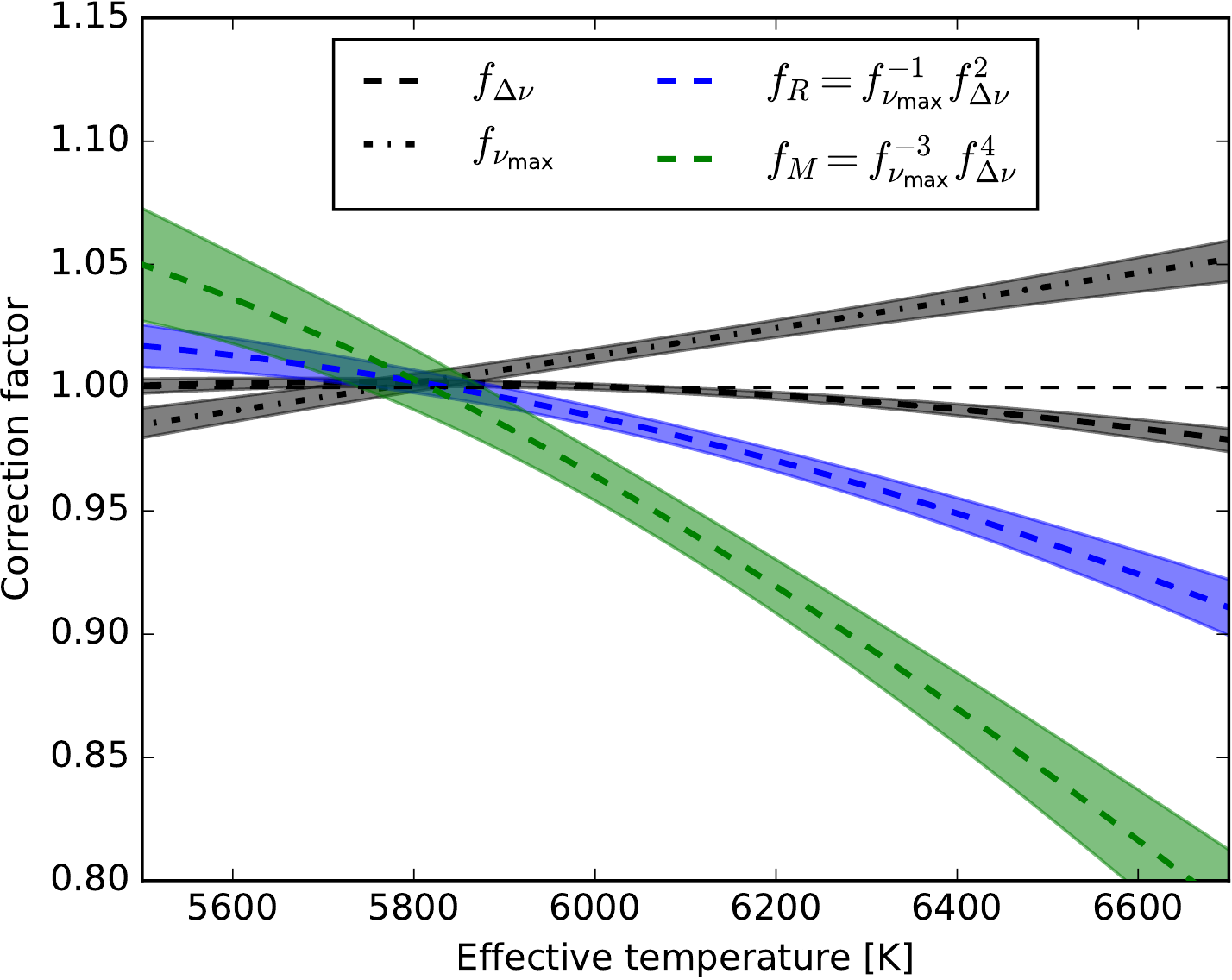}
  \caption{Corrections to $\Delta\nu$ (black dashed), $\nu_{\mathrm{max}}$ (black
           dot-dashed), and the resulting corrections to the scaling relations for the
           radius (blue) and mass (green).}
  \label{fig:scaling_corrections}
\end{figure}

\autoref{fig:scaling_corrections} shows the corrections obtained for $\Delta\nu$ and
$\nu_{\mathrm{max}}$ as well as the corrections they imply for the scaling relations
giving the mass and radius.
The correction to $\nu_{\mathrm{max}}$ is greater than the correction to $\Delta\nu$ at
all temperatures; however, due to the greater exponent on $\Delta\nu$ in the scaling
relations, the two corrections have an almost equal impact on the stellar radius and mass
at the highest temperatures.
These corrections imply that the scaling relations overestimate the radii and masses of
main-sequence stars by $(8\pm1.5)\%$ and $(22\pm4)\%$, respectively, at a temperature
of 6600~K.
However, they also show that for main-sequence stars with temperatures below 6400~K
(which is 85\% of the stars in this study) radii and masses from the scaling relations
are accurate to within 5\% and 13\%, respectively.
Due mostly to the large scatter around $f_{\nu_{\mathrm{max}}}$, these numbers represent
averages for the frequency sample and do not apply to each of the stars individually.

\citet{Coelho2015n} performed a test of the $\nu_{\mathrm{max}}$ scaling relation using
the same method applied here, but to about 400 of the {\it Kepler} dwarfs with fits to
$\Delta\nu_{\mathrm{fit}}$ instead of frequency ratios.
They only found a constant offset from the scaling relation of a few percent and no
significant trend with temperature.
Even though the results presented here are based on frequency ratios which give more
precise (and likely accurate) stellar parameters than $\Delta\nu_{\mathrm{fit}}$, the
sample size is also considerably smaller, and it is not unthinkable that the trend
disappears with more observations.
In fact, if all data points below a temperature of 5800~K are removed from
\autoref{fig:scaling_dnu_numax_corr}, the linear fit to $\nu_{\mathrm{max}}$ is instead
completely flat suggesting a constant offset of about 3\% independent of the temperature.
An updated analysis in the future using individual frequencies for a larger sample of
main-sequence stars would help shed more light on this issue.
There is also the potential of including subgiant stars with individual frequencies, in
order to extend this analysis to later evolutionary stages, but detailed frequency
fitting of subgiants still takes a lot of effort due to their complex oscillation spectra
including mixed modes (see e.g. \citet{Grundahl2017,Brandao2011b}).

It is interesting that none of the results has shown any indications that the scaling
relations need a metallicity-dependent correction.
As mentioned previously, \citet{Viani2017} pointed out that the approximation of the
acoustic cutoff frequency, which is used to derive the scaling relation for
$\nu_{\mathrm{max}}$, introduces a systematic offset in radius that depends on metallicity
due to the omission of a factor related to the mean molecular weight.
In fact, the offset should be ${\sim}6\%$ in radius between the lowest and highest
metallicities of the samples considered in this study.
This is at the same level as the observed temperature dependence and it should be
detectable with the current precision of the parallaxes.
However, no such trend is seen in the parallax comparisons for either of the samples or
in the direct comparison between $\nu_{\mathrm{max}}$ from observations and the scaling
relations.
It is possible that inaccuracies in the relation between $\nu_{\mathrm{max}}$ and
$\nu_{\mathrm{ac}}$ somehow makes up for the inaccuracies in $\nu_{\mathrm{ac}}$, but
until reliable theoretical predictions of $\nu_{\mathrm{max}}$ can be made, it is not
possible to know for sure.
Still, it would be an interesting continuation of this work to include the proposed
metallicity correction to $\nu_{\mathrm{max}}$ in the grids of stellar models and test its
impact on the obtained radii.

Finally, it is important to keep in mind the assumptions that have led to the proposed
corrections to the scaling relations.
They depend directly on the actual accuracy of the radii and masses given by the ratio
fits which have been used as reference values.
While the corrections remove the observed trend with temperature in the parallax
comparison, there is still a constant offset of 3\% which remains unexplained, and the
corrections should be applied with this caveat in mind.

\section{Conclusions}
\label{sec:conclusions}
The main results of comparing asteroseismic and {\it Gaia} parallaxes can be summed up as
follows.

\begin{description}
  \item Parallaxes based on the fits to frequency ratios are, on average, about 3\%
        higher than the {\it Gaia} parallaxes which leads to an absolute offset which increases
        with parallax. The offset does not depend on the adopted temperatures or the
        method used to calculate distances.
  \item For parallaxes based on radii from the scaling relations, the temperature scale
        does affect the comparison.
        Using IRFM temperatures and angular diameters instead of spectroscopic
        temperatures and the distance modulus decreases the offset by about 2\%,
        bringing the parallaxes into agreement with {\it Gaia} on average.
        However, the offset becomes temperature dependent.
        Within about 200~K of the solar temperature, the scaling relations
        agree on the offset found from the ratio fits, and at both lower and higher
        temperatures the offset decreases.
  \item None of the results indicated any dependence of the offset on metallicity.
\end{description}

Due to the indirect nature of the parallax comparison, it is difficult to convert it
directly into an overall accuracy of the seismic radii.
In the end, the results indicate that the scaling relation for the radius is accurate to
within 5\% for main-sequence and subgiant stars.
No tighter constraints can be placed on the analysis until more precise and accurate
parallaxes become available from future {\it Gaia} data releases.
What is more interesting then is the fact that the fractional offset between seismic
parallaxes based on the scaling relations and {\it Gaia} parallaxes change as a function of
temperature.
Based on the good agreement between the scaling relations and the ratio fits at the solar
temperature, it seems most likely that the deviation between the offsets at other
temperatures reflect inaccuracies in the scaling relations, rather than improved
accuracy.

By adopting the stellar parameters obtained from ratio fits as reference values, we have
defined corrections to the scaling relations for $\Delta\nu$ and $\nu_{\mathrm{max}}$.
It is important to note that the validity of these corrections depend on the actual
accuracy of the reference values.
In adopting the radii from the ratio fits as reference values we assume that the 3\%
parallax offset is due to a systematic error in the {\it Gaia} parallaxes or in the
bolometric corrections and extinctions used to convert the seismic radii into distances
(or perhaps a combination of these).
The corrections are found to mainly be functions of effective temperature, and no
significant dependence on metallicity is found despite the fact that the scaling relation
for the acoustic cut-off frequency needs to be corrected for the star's mean molecular
weight.
This may be due to inaccuracies in the scaling between $\nu_{\mathrm{max}}$ and
$\nu_{\mathrm{ac}}$, and a test of the effect of adding a mean molecular weight term to
the scaling relation in the grid-based method would help reveal its impact on the stellar
parameters.

The obtained corrections imply that the scaling relations systematically overestimate
both radii and masses for main-sequence stars at super-solar temperatures.
For stars with temperatures below 6400~K, the deviations from the scaling relations stay
below a level of 5\% in radius and 13\% in mass.
The correction to $\nu_{\mathrm{max}}$ is questionable due to a large scatter
around the relation, and the fact that the correlation with temperature is very weak if
the low-temperature stars are excluded.
An extension of the analysis to later evolutionary stages by the inclusion of subgiants
would help shed light on the validity of the proposed linear correlation with
temperature.

\section*{Acknowledgements}
Funding for the Stellar Astrophysics Centre is provided by The Danish National Research
Foundation (Grant agreement No.~DNRF106). V.S.A. acknowledges support from VILLUM FONDEN
(research grant 10118).
L.C gratefully acknowledges support from the Australian Research Council through
Discovery Program DP150100250 and Future Fellowship FT160100402.
This work has made use of data from the European Space Agency (ESA)
mission {\it Gaia} (\url{https://www.cosmos.esa.int/gaia}), processed by
the {\it Gaia} Data Processing and Analysis Consortium (DPAC,
\url{https://www.cosmos.esa.int/web/gaia/dpac/consortium}). Funding
for the DPAC has been provided by national institutions, in particular
the institutions participating in the {\it Gaia} Multilateral Agreement.
This publication makes use of data products from the Two Micron All Sky Survey,
which is a joint project of the University of Massachusetts and the Infrared
Processing and Analysis Center/California Institute of Technology, funded by the
National Aeronautics and Space Administration and the National Science Foundation.



\bibliographystyle{mnras}
\bibliography{references} 




%
%


\bsp	
\label{lastpage}
\end{document}